\documentclass[aps,prm,reprint,showpacs,superscriptaddress]{revtex4-1}
\usepackage{graphicx}
\usepackage{dcolumn}
\usepackage{bm}
\usepackage[dvipdfm]{hyperref}
\hypersetup{colorlinks=true,citecolor=blue,urlcolor=blue,linkcolor=blue}
\usepackage{textcomp}
\begin{document}

\title{Slow vortex creep induced by strong grain boundary pinning in advanced Ba122 superconducting tapes}

\author{Chiheng Dong}
\affiliation{Key Laboratory of Applied Superconductivity, Institute of Electrical Engineering, Chinese Academy of Sciences, Beijing 100190, People's Republic of China}

\author{He Huang}
\affiliation{Key Laboratory of Applied Superconductivity, Institute of Electrical Engineering, Chinese Academy of Sciences, Beijing 100190, People's Republic of China}
\affiliation{University of Chinese Academy of Sciences, Beijing 100049, People's Republic of China}

\author{Yanwei Ma}
\email{ywma@mail.iee.ac.cn}%
\affiliation{Key Laboratory of Applied Superconductivity, Institute of Electrical Engineering, Chinese Academy of Sciences, Beijing 100190, People's Republic of China
}%
\affiliation{University of Chinese Academy of Sciences, Beijing 100049, People's Republic of China}

\date{\today}

\begin{abstract}
We report the temperature, magnetic field and time dependences of magnetization in advanced Ba122 superconducting tapes. The sample exhibits a peculiar vortex creep behavior. Below 10 K, the normalized magnetization relaxation rate S=dln(-M)/dln(\textit{t}) shows a temperature insensitive plateau with a value comparable to that of the low temperature superconductors, which can be explained within the framework of the collective creep theory. It then enters into a second collective creep regime when the temperature increases. Interestingly, the relaxation rate below 20 K tends to saturate with the increasing field. However, it changes to a power law dependence on field at a higher temperature. A vortex phase diagram composed of the collective and the plastic creep regions is concluded. Benefit from the strong grain boundary pinning, the advanced Ba122 superconducting tape has promising potential to be applied not only in liquid helium but also in liquid hydrogen or at the temperature accessible with cryocoolers.
\end{abstract}

\maketitle

Understanding the behavior of vortex matter in a superconductor is vitally important for both basic physics and technological applications. Vortex creep caused by thermal fluctuation introduces measurable dissipation and a reduction of the maximum loss-less current. The scale of the thermal fluctuation is usually parameterized by the dimensionless Ginzburg number Gi$\propto$$\gamma$$^2$T$_c$$^2$$\lambda$$^4$/$\xi$$^2$, where $\gamma$ is the anisotropy parameter, T$_c$ is the superconducting transition temperature, $\lambda$ is the penetration depth and $\xi$ is the coherence length. Due to high T$_c$, small $\xi$ and large $\gamma$, the thermal fluctuation in cuprates is significant, giving rise to the so-called giant vortex creep \cite{yeshurun_magnetic_1996}. It is thus a major task to reduce the detrimental effect of vortex motion in cuprates by incorporating effective pinning centers. Iron-based superconductors (IBSC), on the contrary, have larger $\xi$, lower $\gamma$, and consequently smaller Gi than that of the cuprates \cite{kwok_vortices_2016}. In combination with their high upper critical field H$_{c2}$ and moderate T$_c$ \cite{gurevich_use_2011}, IBSC are considered as potential candidates in large-scale application of superconductivity.

After ten years of research and design since the discovery of IBSC, the (Ba/Sr)$_{0.6}$K$_{0.4}$Fe$_2$As$_2$ superconducting tapes fabricated by the powder in tube (PIT) method \cite{ma_progress_2012,hosono_recent_2018} are now the prevalent materials for applied research. Up to now, the critical current density $J_c$(4.2 K, 10 T) of the (Ba/Sr)$_{0.6}$K$_{0.4}$Fe$_2$As$_2$ tape has already surpassed the practical application level \cite{zhang_realization_2014}. Recently, we optimized the hot pressing method and enhance the transport $J_c$ to 1.5$\times$10$^5$ A/cm$^2$ at 4.2 K and 10 T \cite{huang_high_2018}. It even retains a J$_c$ of 5.4$\times$10$^4$ A/cm$^2$ at 20 K and 5 T. It was found that the grain size is only 0.5-1 $\mu$m, much smaller than that of the Sr$_{0.6}$K$_{0.4}$Fe$_2$As$_2$ counterpart \cite{lin_hot_2014} whose grains are 4-7 $\mu$m large. Small grains give rise to increased grain boundary density and consequently strong vortex pinning strength\cite{dong_vortex_2016}. It will probably make a difference to the vortex motion behavior.

In this paper, we measure the magnetization of the same Ba122 superconducting tape studied in Ref.\cite{huang_high_2018} via a vibrating sample magnetometer equipped on a Physical Property Measurement System (PPMS-9). Time dependence of magnetization was obtained over 1 hour with the field perpendicular to the tape surface. The ramp rate of the magnet is set to 160 Oe/s during the measurement. The dynamic magnetic relaxation method is not used here because the difference of the magnetization under different magnet ramp rates is quite small to be detected by our equipment.

The temperature dependence of magnetization is depicted in Fig.1(a). The superconducting transition temperature is $\sim$38 K. After the transition, a strong diamagnetic signal with little temperature dependence emerges in the ZFC curve. The superconducting volume fraction is nearly 100 \%. The FC branch shows a very small Meissner vortex expulsion. Fig.1(b) shows an isothermal M-H curve at 4 K with field perpendicular to the tape surface. Interestingly, we observe irregular discontinuities known as vortex jumps in the M-H loop below 1 T. This magnetic instability retains up to 6 K and disappears afterwards. The vortex jumps can be widely observed in many type-\uppercase\expandafter{\romannumeral2} superconductors, such as MgB$_2$ \cite{romero-salazar_flux_2007}, cuprates \cite{nabialek_magnetic_2003} and IBSC \cite{wang_very_2010,pramanik_flux_2013}. Generally, it appears in the M-H loop of a large sample carrying a high $J_c$ under a fast ramping magnetic field \cite{wang_very_2010}. The large hysteresis loop indicates that there is a large global current across the sample. The calculated magnetic critical current density $J_c^{mag}$ based on the Bean model is shown in Fig.1(e). At 9 T and 4 K, the $J_c^{mag}$ is 1.55$\times$10$^5$ A/cm$^2$, very close to the transport $J_c$ \cite{huang_high_2018}.

\begin{figure}
  \includegraphics[width=8cm]{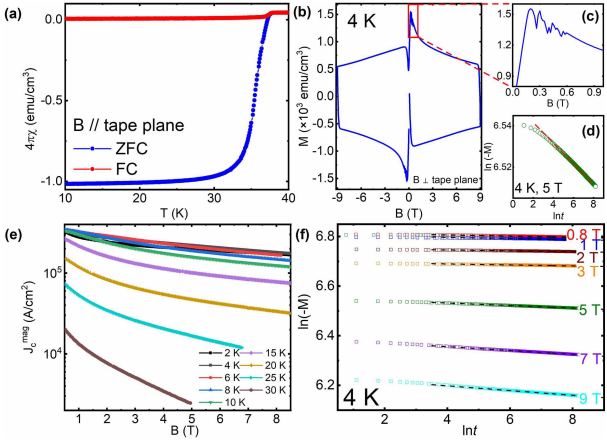}\\
  \caption{(a) Temperature dependence of magnetization with ZFC and FC procedures. (b) Isothermal M-H loop at 4 K. (c) A clear vortex jump can be observed at low field. (d) M-\textit{t} curve at 4 K and 5 T shows two different relaxation stages. (e) Field dependence of J$_c^{mag}$ between 2 K and 30 K. (f) Time dependence of magnetization on double logarithmic scales at 4 K. The dashed lines are linear fit to the curves.}
  \label{Figure1}
\end{figure}

Fig.1(f) shows the time dependence of magnetization on double logarithmic scales at 4 K. There is only 1 \% and 6.4 \% loss of magnetization over 1 hour at 1 T and 9 T, respectively. It is therefore reasonable to see a rather robust field dependence of transport $J_c$ at 4.2 K. We find two stages of vortex relaxation. As shown in Fig.1(d), the initial nonlogarithmic stage with a slower relaxation rate is due to a transient effect \cite{gurevich_time_1993} which is commonly observed in M(\textit{t}) curves \cite{salem-sugui_fishtail_2011}. In the second stage, the magnetization depends logarithmically on time, consistent with the model of thermally activated vortex motion. The normalized relaxation rate S=dln(-M)/dln\textit{t} can thus be obtained from the slope of the ln(-M)-ln\textit{t} curve, as shown with the dashed lines in Fig.1(d).

Fig.2(a) depicts the normalized magnetization relaxation rate as a function of magnetic field. The relaxation rate exhibits nonmonotonic field dependence. At 2 K, the relaxation rate begins to increase with the field, reaches a peak at 0.8 T and decreases afterwards. Then it gets to a minimum at 2 T and finally increases with the field. The curves at 4 K and 6 K are quite similar to that of 2 K except that the position of the minimum shifts to a lower temperature. In addition, the relaxation rate at 6 K increases slowly and tends to saturate at high field. The saturation becomes more obvious with the increasing temperature. The saturation value at 6 K, 8 K, 10 K and 15 K is 0.014, 0.02, 0.023 and 0.04, respectively. Generally, the S(H) curve of a high temperature superconductor increases monotonically at high field, \textit{e.g.} MgB$_2$ \cite{wen_magnetic_2001} and YBCO crystals \cite{yeshurun_flux_1989}. As a result, the effective pinning energy U$_{eff}$(H)=T/S(H), which is inversely proportional to the relaxation rate at a definite temperature \cite{yeshurun_magnetic_1996}, will quickly decrease with the field. On the contrary, the S(H) in our case varies slowly below 20 K, leading to a robust U$_{eff}$(H) at high field. This is beneficial to high field application. However, the situation has changed when the temperature is raised above 20 K, implying a crossover to a different vortex creep regime. The relaxation rates at 20 K, 25 K and 30 K increase quickly with the field. We use a power law equation S(H)$\propto$H$^n$ to fit the curve, where \textit{n} is the power exponent. As shown by the black dashed lines in Fig.2(a), the fitted \textit{n} values for 20 K, 22.5 K and 25 K is 1.8, 2 and 3, respectively. This power law relation is also observed in the cuprates \cite{yeshurun_magnetic_1996} and the IBSC. One example is shown by the dashed line in Fig.4.

\begin{figure}
  \includegraphics[width=8cm]{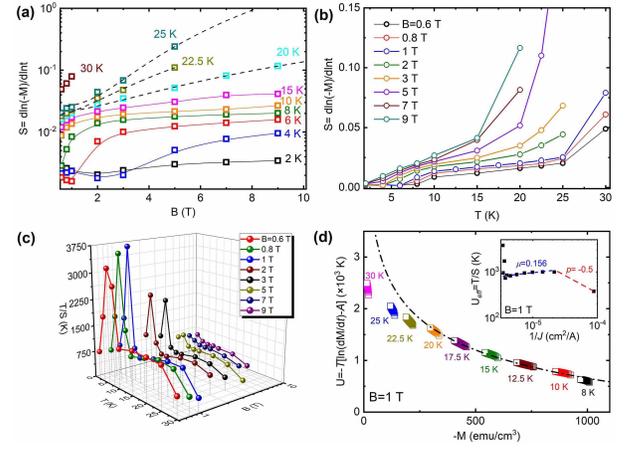}\\
  \caption{(a) Normalized magnetization relaxation rate as a function of magnetic field. (b) Temperature dependence of relaxation rate S. (c) Temperature dependence of T/S. (d) Magnetization dependence of the scaled activation energy U at 1 T obtained by the Maley's method. The inset is the effective pining energy U$_{eff}$ as a function of 1/\textit{J} at 1 T.}
  \label{Figure2}
\end{figure}

The temperature dependence of relaxation rates are concluded in Fig.2(b). At 0.6 T, the S(T) curve shows a plateau like behavior below 10 K, which can be explained within the framework of the collective creep theory. The plateau area shrinks with the increasing field, and disappears after 5 T. We find that the plateau value is independent of the field and remains at 0.002. This relaxation rate is one order of magnitude smaller than that of the cuprates \cite{civale_defect_1990}, and it is comparable to that of the NbTi wire, of which the creep rate at 2.5 K is 0.00235 and 0.00432 for 0.3 T and 1 T, respectively \cite{eley_universal_2017}. When the temperature increases, the relaxation rate at 0.6 T undergoes a sudden jump at 10 K, gradually increases thereafter, and finally diverges at 25 K. It implies a successive transformation of vortex creep regime. The S(T) curves below 5 T follow the similar trend. While at higher field, the S(T) curves show linear temperature dependence, and the divergence shifts to lower temperature.

The collective creep theory supposes a diverging pinning energy U(\textit{J}) when the current density diminishes \cite{feigelman_flux_1991}. To guarantee this, a widely used interpolation that covers all known functional forms of U(\textit{J}) is proposed: U(J,T,B)=$\frac{U_0(T,B)}{\mu(T,B)}[(\frac{J_c(T,B)}{J(T,B)})^{\mu(T,B)}-1]$, where U$_0$ is the characteristic pinning energy, $\mu$ is the glassy exponent determined by the bundle size and vortex lattice elasticity, $J_c$ is the critical current density in the absence of vortex creep \cite{malozemoff_flux_1991}. Generally, a positive $\mu$ corresponds to the elastic creep regime, while a negative $\mu$ corresponds to the plastic creep regime \cite{wen_field_1997}. Combining with the general formula of U(\textit{J})=Tln($t/t_0$), where $t_0$ is a macroscopic characteristic time depending on the sample size and shape \cite{yeshurun_magnetic_1996}, one can obtain: $J(T,t)$=$J_c$/[1+($\mu$T/$U_0$)ln($t/t_0$)]$^{1/\mu}$. The normalized relaxation rate S=dlnM/dln\textit{t}=dlnJ/dln\textit{t} can be expressed as: $S=T/(U_0+\mu Tln(t/t_0))$. The effective pinning energy U$_{eff}$ is:
\begin{equation}
U_{eff}=T/S=U_0+\mu Tln(t/t_0)=U_0(J_c/J)^\mu.
\label{eq:1}
\end{equation}

Fig.2(c) shows the T/S-T curves at different fields. Interestingly, there are two peaks below 7 T. The first prominent peak is below 10 K. With the increasing field, it gradually shrinks, shifts to a lower temperature and finally moves beyond the temperature limit of our PPMS. The second broad peak is at a higher temperature. Different from most cases of the high-temperature superconductors whose T/S-T curves only present one peak, the situation here is more analogous to the Tl$_2$Ba$_2$CaCu$_2$O$_8$ film, which can be explained as the crossover between different collective pinning regimes \cite{wen_flux_2001}. It is proved that a positive slope of the T/S-T curve corresponds to a positive $\mu$ and the elastic vortex motion, while a negative slope corresponds to a negative $\mu$ and the plastic vortex motion \cite{shen_flux_2010}. Based on this theory, our results suggest that there are two different collective creep regimes. The first one is located below 10 K where a prominent peak appears. The second is in the intermediate temperature region below the second broad peak. While at a higher temperature, the slope changes to negative and the vortex creep become plastic.

For a closer inspection on the vortex creep behavior in the intermediate temperature region, we study the relation between the pinning energy and the critical current density. Firstly, we depict the effective pinning energy as a function of 1/\textit{J} at 1 T, as shown in the inset of Fig.2(d). According to Equ.\ref{eq:1}, the slope in the double-logarithmic plot of U$_{eff}$ \textit{vs.} 1/\textit{J} gives the value of the glassy exponent. In the intermediate temperature region, the evaluated value of $\mu$ is $\sim$0.156, close to the value of single vortex creep \cite{feigelman_theory_1989}. With the decreasing \textit{J}, the slope becomes negative with a value of \textit{p}=-0.5, corresponding to the plastic creep regime. This result is further corroborated by the analysis of the U(\textit{J}) relation via the Maley's method \cite{maley_dependence_1990}. It proposes a general equation $U=T[ln(dM/dt)-A]$ to scale the data measured at different temperatures, where \textit{A} is a time-independent constant associated with the average hopping velocity. The main panel of Fig.2(d) shows the U-M curves between 8 K and 30 K. The curves below 6 K are not shown here because of decreased M value resulting from the flux jumps at 1 T. We adjust the constant \textit{A} and find that the U(M) between 8 K and 20 K fall on a smooth curve. We fit the curve using U=U$_0$[(J$_c$/J)$^\mu$-1]/$\mu$, as shown by the dashed line. The fitted glassy exponent is $\mu$=0.15, consistent with the value evaluated from the U$_{eff}$(1/\textit{J}) curve. The deviation of the data from the fitting line above 20 K is reasonable since the vortex creep become plastic.

\begin{figure}
  \includegraphics[width=8cm]{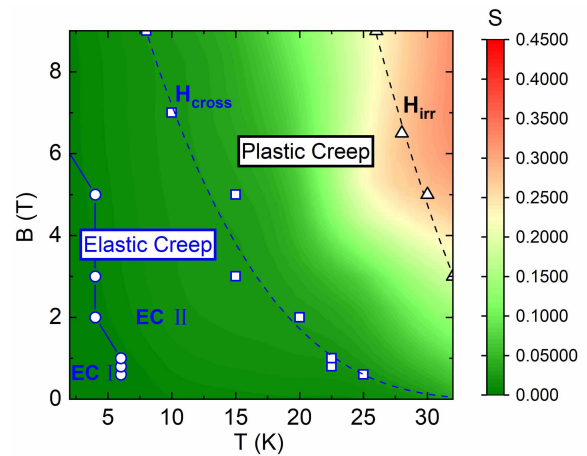}\\
  \caption{Vortex phase diagram of the Ba122 superconducting tape. The contours with different colors correspond to the relaxation rates with different values.}
  \label{Figure3}
\end{figure}

With the data in hand, we conclude a vortex phase diagram as schematized in Fig.3. The black triangle denotes the irreversible field H$_{irr}$ derived from the M-H loops. The H$_{cross}$ corresponding to the second broad peak in the T/S-T curve is shown with the blue square. It demarcates the regions of elastic and plastic vortex creep. The blue circle derived from the first prominent T/S-T peak correspond to the crossover between two different elastic creep regimes, EC \uppercase\expandafter{\romannumeral1} and EC \uppercase\expandafter{\romannumeral2}. The temperature dependence of the characteristic fields can be well fitted by the expression $H_c=H_c(0 K)(1-T/T_c)^n$, as shown by the dashed lines. For the irreversible field, H$_{irr}$(0 K)=53 T, \textit{n}=1.55. For the crossover filed, H$_{cross}$(0 K)=19.2 T, \textit{n}=3.23. At liquid helium temperature, the relaxation rate remains below 0.01 up to 9 T, indicating a promising application at high magnetic field. It is proved that the grain boundary in the iron-based superconducting wires and tapes is the main pinning source \cite{dong_vortex_2016}. Small grains correspond to large grain boundary pinning force \cite{martinez_flux_2007}, large operable field range and large critical current at high field \cite{hecher_small_2016}. In our case, the grains are only 0.5-1 $\mu$m large \cite{huang_high_2018}, which are 4-10 times smaller than that of the Sr122 tape \cite{lin_hot_2014}, which may lead to a wider elastic creep area \cite{dong_vortex_2016}. Above the H$_{cross}$, the thermally activated vortex creep comes from the sliding of dislocations of vortex lattice rather than jumps of vortex bundles. When the magnetic field approaches the H$_{irr}$, the relaxation rate steepens, the vortices flow freely and the loss-less current disappears.

\begin{figure}
  \includegraphics[width=8cm]{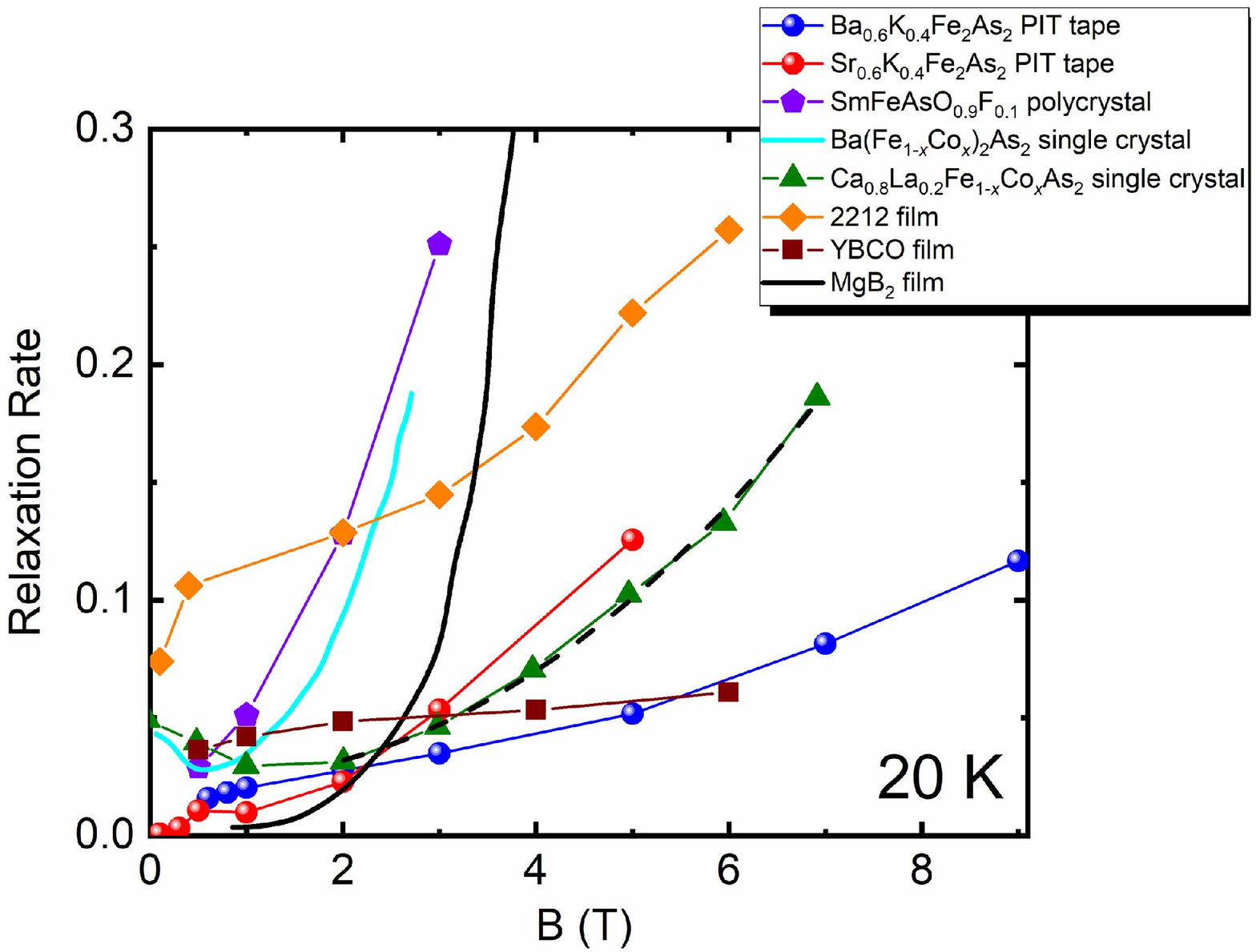}\\
  \caption{Field dependence of relaxation rate at 20 K for the the Ba$_{0.6}$K$_{0.4}$Fe$_2$As$_2$ PIT tape in this work, the Sr$_{0.6}$K$_{0.4}$Fe$_2$As$_2$ PIT tape \cite{dong_vortex_2016}, the SmFeAsO$_{0.9}$F$_{0.1}$ polycrystal \cite{yang_magnetization_2008}, the Ba(Fe$_{1-x}$Co$_x$)$_2$As$_2$ single crystal \cite{shen_flux_2010}, the Ca$_{0.8}$La$_{0.2}$Fe$_{1-x}$Co$_x$As$_2$ single crystal \cite{zhou_second_2016}, the 2212 film \cite{wen_field_1997}, the YBCO film \cite{wen_critical_1995} and the MgB$_2$ film \cite{wen_magnetic_2001}. The black dashed line is fitted curves using S(H)$\propto$H$^n$}
  \label{Figure4}
\end{figure}

To investigate the application prospect at liquid hydrogen temperature, we compare the field dependence of the vortex creep rates of several high temperature superconductors that have probable application at 20 K. As shown in Fig.4, the relaxation rate of MgB$_2$ quickly increases with the field due to a small irreversible field H$_{irr}$$\sim$4 T at 20 K \cite{wen_magnetic_2001}. For the 2212 system, large anisotropic parameter $\gamma$$\sim$50 and small coherence length $\xi$ render a small pinning energy $\sim$(H$_{c2}$/8$\pi$)($\xi$$^3$/$\gamma$) and a large scale of thermal fluctuation as revealed by a large Gi$\sim$1. As a result, the vortex creeps fast even at low field \cite{wen_field_1997}. The situation is ameliorated for the YBCO which has a smaller $\gamma$$\sim$8, and consequently a smaller Gi$\sim$10$^{-2}$ as well as a larger pinning energy. The IBSC share a moderate Gi ($\sim$10$^{-4}$ for the 122 system) that is intermediate between the low temperature superconductors (Gi$\sim$10$^{-8}$) and the cuprates. It is thus believed that IBSC should have creep rates between those of the low temperature superconductors and the cuprates. However, the vortex dynamics of IBSC is quite different from one another because of the diverse pinning landscapes. As shown in Fig.4, the high-field creep rate of the hot-pressed Ba$_{0.6}$K$_{0.4}$Fe$_2$As$_2$ tape is smaller than that of the SmFeAsO$_{0.9}$F$_{0.1}$ polycrystals, the hot-pressed Sr122 tapes, and even the optimally doped Ba(Fe$_{1-x}$Co$_x$)$_2$As$_2$ and Ca$_{0.8}$La$_{0.2}$Fe$_{1-x}$Co$_x$As$_2$ single crystals. Combining with its high upper critical field H$_{c2}$(20 K)$\sim$45 T \cite{gurevich_use_2011} and large transport $J_c$(5 T, 20 K)$\sim$5.4$\times$10$^4$ A/cm$^2$ \cite{huang_high_2018}, we suggest that the Ba122 tape has great potential to be applied in high field magnets operated with liquid hydrogen or cryocoolers.

In summary, the advanced Ba122 superconducting tape with high critical current density exhibits slow vortex creep below 10 K, of which the value is comparable to that of the low temperature superconductors. It is found that the vortex creep behavior in the low and intermediate temperature region can be well explained by the model of collective creep. Interestingly, the S(H) curves below 20 K saturate at high field which is beneficial to high field applications. Our studies indicate that the advanced Ba122 superconducting tape has promising potential to be applied in high field magnets operated not only with liquid helium, but also with liquid hydrogen or cryocoolers.

This work is partially supported by the National Natural Science Foundation of China (Grant Nos. 51402292, 51677179), the International Partnership Program of Chinese Academy of Sciences (grant Nos. GJHZ1775 and 182111KYSB20160014), the Key Research Program of Frontier Sciences, CAS (QYZDJ-SSW-JSC026), Strategic Priority Research Program of Chinese Academy of Sciences (Grant No. XDB25000000).

\bibliography{Reference}

\end{document}